\documentclass{elsart}
\usepackage{epsfig}
\begin{document}
\begin{frontmatter}
\title{Energy distribution analysis of the wavepacket simulations of
  CH${}_4$ and CD${}_4$ scattering}
\author{R.\ Milot and A.\ P.\ J.\ Jansen}
\address{Schuit Institute of Catalysis, ST/SKA, Eindhoven University of
  Technology\\ P.O. Box 513, NL-5600 MB Eindhoven, The Netherlands.\\
  E-mail: tgakrm@chem.tue.nl, Tel.:+31-40-2472189,  Fax:+31-40-2455054} 
\date{\today}
\maketitle

\begin{abstract}

The isotope effect in the scattering of methane is studied by
wavepacket simulations of oriented CH${}_4$ and CD${}_4$ molecules from
a flat surface including all nine internal vibrations. At a
translational energy up to 96 kJ/mol we find that the scattering is
still predominantly elastic, but less so for CD${}_4$.
Energy distribution analysis of the kinetic energy per mode and the
potential energy surface terms, when the molecule hits the surface, are
used in combination with vibrational excitations and the corresponding
deformation. They indicate that the orientation with three bonds
pointing towards the surface is mostly responsible for the isotope
effect in the methane dissociation.  

\end{abstract}

\end{frontmatter}
keywords: Computer simulations, Models of surface chemical reaction,
Alkanes, Low index single crystal surfaces

\section{Introduction}

The dissociation of methane on transition metals is an important
reaction in catalysis. It is the rate limiting step in steam reforming
to produce syngas.\cite{hoo80} It is also prototypical for C--H
activation in other processes.  A large number of molecular beam
experiments in which the dissociation energy was measured as a function
of translational energy have already been done on this
system.\cite{ret85,ret86,lee87,hol95,lar99,lun89,walker99,lee86,cey87,bec89,hol96,xxx11,xxx13,val96a,val96b,ver93,ver94,ver95,see97,see97b,juur99}
These experiments have contributed much to our understanding of the
mechanism of the dissociation. Some of them observed that vibrationally
hot ${\rm CH}_4$ dissociates more readily than cold ${\rm CH}_4$, with
the energy in the internal vibrations being about as effective as the
translational energy in inducing
dissociation.\cite{ret85,ret86,lee87,hol95,lar99,lun89,walker99} A more
detailed assessment of the importance of the internal vibrations could
not be made, because of the large number of internal vibrations.  A
recent molecular beam experiment with laser excitation of the $\nu_3$
mode succeeded in measuring a dramatical enhancement of the dissociation
on a Ni(100) surface, but it is still much too low to account for the
vibrational activation observed in previous studies and indicates that
other vibrationally excited modes contribute significantly to the
reactivity of thermal samples.\cite{juur99}

Wavepacket simulations are being used more and more to study the
dynamics of this kind of molecule surface reactions. The published
wavepacket simulations on the methane dissociation reaction on
transition metals have treated the methane molecule always as a diatomic
up to now.\cite{lun91,lun92,lun95,jan95,car98} Besides the C--H bond and
molecule surface distance, a combination of other coordinates were
included, like (multiple) rotations and some lattice motion. None of
them have looked at the role of the internal vibrations. Various
theoretical studies have obtained reaction pathways and barriers for
dissociation by DFT
calculations,\cite{yan92,bur93a,bur93b,bur94,bur95a,kra96,au97,liao97,au99,beng99}
but they cannot explain the role of the vibrational modes in the
reaction dynamics either.

A nice way to study reaction dynamics is the use of isotopes. The most
recent wavepacket simulation on the dissociation probability of CH${}_4$
and CD${}_4$ showed a semiquantitative agreement with the molecular beam
experiments of Ref.\cite{hol95}, except for the isotope effect 
and the extracted vibrational efficacy.\cite{car98} The molecular beam
study with laser excitation of the $\nu_3$ asymmetrical stretch mode
shows that the incorrect vibrational efficacy is caused by the
assumptions in the fit procedure that both stretch modes behaves
identical.\cite{juur99} One of the possible explanation of the
incorrect isotope effect can be the role played by the non-included
intramolecular vibrations. 

In a previous paper we reported on wavepacket simulations that we have
done to determine which and to what extent internal vibrations are
important for the dissociation of CH${}_4$.\cite{mil98} We were not able
yet to simulate the dissociation including all internal vibrations.
Instead we simulated the scattering of methane, for which all internal
vibrations can be included, and used the results to deduce consequences
for the dissociation. We used model potential energy surfaces (PESs)
that have been developed with Ni(111) in mind, but our results should
hold for other surfaces as well.  At a translational energy up to 96
kJ/mol we found that the scattering is almost completely elastic.
Vibrational excitations when the molecule hits the surface and the
corresponding deformation depend on generic features of the potential
energy surface. In particular, our simulations indicate that for methane
to dissociate the interaction of the molecule with the surface should
lead to an elongated equilibrium C--H bond length close to the surface.

We have been using the multiconfigurational time-dependent Hartree
(MCTDH) method for our wavepacket simulation, because it can deal with
a large number of degrees of freedom and with large
grids.\cite{man92,jan93} This method has been applied successfully to
gas phase reactions and reactions at
surfaces.\cite{man92b,man93,ham94,fan94,fan95,fan95b,liu95,cap95,eha96,wor96,jac96,man96,mat96,ger97,bec97,jac98a,jac98b,wor98,mat99,jac99}

In this paper we report wavepacket simulations of CD${}_4$ scattering
including all internal vibrations for fixed orientations, performed on
the same model PESs as in our previous paper.\cite{mil98} Translational
motion parallel to the surface and all rotational motion was neglected.
No degrees of freedom of the surface were included.  Experiments show
that coupling with these degrees of freedom is dependent on the metal
surface. For example, the observed surface temperature effect are small
on Ni\cite{hol95}, but quite large on Pt\cite{lun89}. As we are only
interested in the role of internal vibrations, we have not included
degrees of freedom of the surface to keep the simulations as simple as
possible.  We will discuss the vibrational excitation and the
deformation of the CD${}_4$ molecule when it hits the surface and
compare it with CH${}_4$. Later on we will look at the energy
distribution of the kinetic energy per mode and the potential energy in
some terms of the PES with the elongated equilibrium bond length close
to the surface for both isotopes.  The transfer of translational kinetic
energy towards vibrational kinetic energy gives an indication about the
dissociation probability, since vibrational kinetic energy helps in
overcoming the dissociation barrier.  It gives a better idea too about
which modes are essential to include in a more accurate wavepacket
simulation of methane dissociation. After that we will discuss the
implications of this for the dissociation and give a summary with some
general conclusions.

\section{Computational details}

\subsection{The Potential Energy Surfaces}

We used for the scattering of CD${}_4$ the same model PESs as we did for
CH${}_4$. Since we expressed the PES in mass-weighted coordinated the
parameters in the PESs for CD${}_4$ differs from CH${}_4$. We will now
give an overview of our model PESs and the corresponding parameters for
CD${}_4$. The parameters of CH${}_4$ for these PESs were already 
given in Ref.~\cite{mil98}, where also more detailed information
about our assumptions and contour plots of some cross-section of the
model PESs can be found.

The PESs we used can all be written as
\begin{equation}
  \label{Vgen}
  V_{\rm total}=V_{\rm intra}+V_{\rm surf},
\end{equation}
where $V_{\rm intra}$ is the intramolecular PES and $V_{\rm surf}$ is the
repulsive interaction with the surface. For the $V_{\rm intra}$ we
looked at four different types of PESs. The $V_{\rm intra}$ include
also for two types changes in the intramolecular potential due to
interactions with the surface.

\subsubsection{A harmonic potential}

The first one is completely harmonic. We have used normal mode
coordinates for the internal vibrations, because these are coupled only
very weakly.
In the harmonic approximation this coupling is even absent so that we
can write $V_{\rm intra}$ as 
\begin{equation}
  \label{Vharm}
  V_{\rm intra}=V_{\rm harm}= {1\over 2}\sum_{i=2}^{10}k_iX_i^2,
\end{equation}
the summation is over the internal vibrations, $X_i$'s are mass-weighted
displacement coordinates and $k_i$ are mass-weighted force constants. (see
Table \ref{tab:gen} for definitions and values); ($X_1$ is the
mass-weighted overall translation along the surface normal).\cite{wil55}
The force constants have been obtained by fitting them on the
experimental vibrational frequencies of CH${}_4$ and
CD${}_4$.\cite{gray79,lee95} 

We have assumed that the repulsive interaction with the surface is only
through the deuterium atoms that point towards the surface. We take the
$z$-axis as the surface normal. In this case the surface PES is given by
\begin{equation}
  \label{gen_Vsurf}
  V_{\rm surf}={A \over{N_D}}\sum_{i=1}^{N_D} e^{-\alpha z_i},
\end{equation}
where $N_D$ is the number of deuteriums that points towards the surface,
$\alpha$=1.0726 atomic units and $A$=6.4127 Hartree. These parameters
are chosen to give the same repulsion as the PES that has been used in
an MCTDH wavepacket simulation of CH${}_4$ dissociation.\cite{jan95} 

\begin{figure}[h]
  \epsfig{file=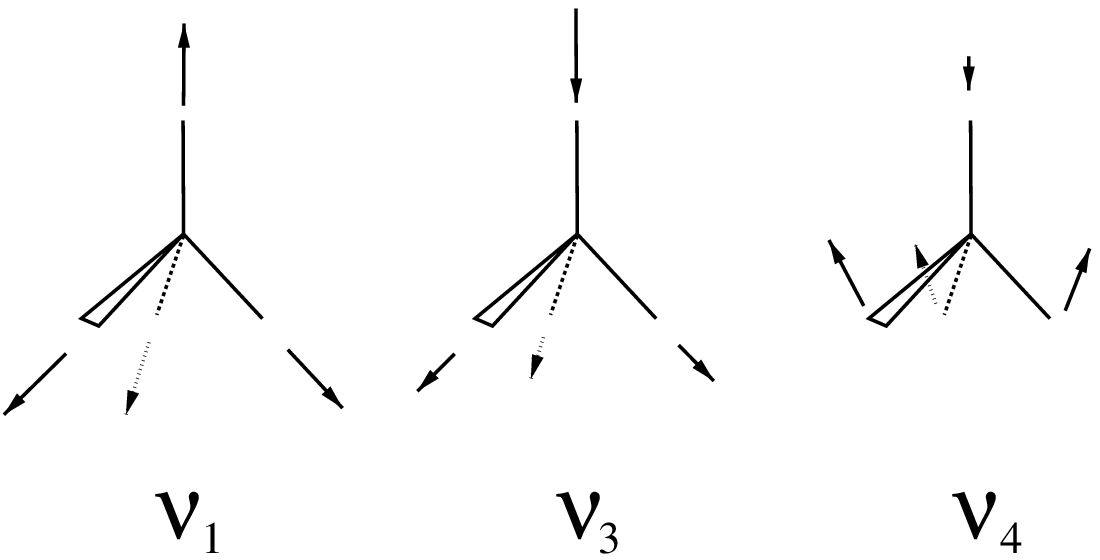, height=7.0cm} 
  \caption{The $a_1$ vibrational normal modes in the C${}_{3v}$ symmetry;
    $\nu_1$ symmetrical stretch (X${}_2$), $\nu_3$ asymmetrical stretch
    (X${}_4$), and $\nu_4$ umbrella (X${}_3$).}
  \label{fig:modes}
\end{figure}

If we write $V_{\rm surf}$ in terms of normal mode coordinates, then we
obtain for one deuterium pointing towards the surface 
\begin{equation}
  \label{vsurfone}
  V_{\rm surf}= A
  e^{-\alpha_1X_1}e^{-\alpha_2X_2}e^{-\alpha_3X_3}e^{-\alpha_4X_4},
\end{equation}
where $A$ as above, and $\alpha$'s as given in Table \ref{tab:alpha}.
$X_2$, $X_3$ and $X_4$ correspond all to $a_1$ modes of the C${}_{3v}$
symmetry (see Fig.~\ref{fig:modes}). There is no coupling between the
modes $X_5$ to $X_{10}$ in the $V_{\rm surf}$ part of the PES, which are
all $e$ modes of the C${}_{3v}$ symmetry.

For two deuteriums we obtain
\begin{eqnarray}
  \label{vsurftwo}
  V_{\rm surf}=A&&e^{-\alpha_1X_1}e^{-\alpha_2X_2}
  e^{-\alpha_3X_3}e^{-\alpha_4X_4}e^{-\alpha_5X_5}\\
  \times{1\over2}\Big[
  &&e^{\beta_{3}X_7}
    e^{-\beta_{3}X_8}e^{\beta_{5}X_9}e^{-\beta_{5}X_{10}}
    \nonumber\\
  +&&e^{-\beta_{3}X_7}
    e^{\beta_{3}X_8}e^{-\beta_{5}X_9}e^{\beta_{5}X_{10}}
    \Big],\nonumber
\end{eqnarray}
with $A$ again as above, $\alpha$'s and $\beta$'s as given in Table
\ref{tab:alpha}.
$X_2$, $X_3$, $X_4$ and $X_5$ correspond all to $a_1$ modes of C${}_{2v}$. 
$X_7$, $X_8$, $X_9$ and $X_{10}$ correspond to $b_1$ and $b_2$ modes of
C${}_{2v}$. $X_6$ corresponds to the $a_2$ mode of C${}_{2v}$ and
has no coupling with the other modes in $V_{\rm surf}$.

For three deuteriums we obtain
\begin{eqnarray}
  \label{vsurfthree}
  V_{\rm surf}=A&&e^{-\alpha_1X_1}e^{-\alpha_2X_2}
  e^{-\alpha_3X_3}e^{-\alpha_4X_4}\\
  \times{1\over 3}\Big[
  &&e^{\beta_{1}X_5}e^{\beta_{2}X_6}e^{-\beta_{3}X_7}
    e^{\beta_{4}X_8}e^{\beta_{5}X_9}e^{-\beta_{6}X_{10}}
    \nonumber\\
  +&&e^{\beta_{1}X_5}e^{-\beta_{2}X_6}e^{-\beta_{3}X_7}
    e^{-\beta_{4}X_8}e^{\beta_{5}X_9}e^{\beta_{6}X_{10}}
    \nonumber\\
  +&&e^{-2\beta_{1}X_5}e^{2\beta_{3}X_7}e^{-2\beta_{5}X_9}
    \Big],\nonumber
\end{eqnarray}
with $A$ again as above, $\alpha$'s and $\beta$'s as given in Table
\ref{tab:alpha}.  $X_2$, $X_3$ and $X_4$ corresponds to $a_1$ modes in
the C${}_{3v}$ symmetry (see Fig.~\ref{fig:modes}).  Because these last
six coordinates correspond to degenerate $e$ modes of the C${}_{3v}$
symmetry, the $\beta$ parameters are not unique.

\subsubsection{An anharmonic intramolecular potential}
\label{sec:morse}

Even though we do not try to describe the dissociation of methane in
this and our previous paper, we do want to determine which internal
vibration might be important for this dissociation. The PES should
at least allow the molecule to partially distort as when
dissociating. The harmonic PES does not do this. A number of
changes have therefor been made. The first is that we have describe the
C--D bond by a Morse PES.
\begin{equation}
  \label{Vmorse_n}
  V_{\rm Morse}=D_e \sum_{i=1}^4\Big[1-e^{-\gamma \Delta r_i}\Big]^2 ,
\end{equation}
where $D_e=0.1828$ Hartree (the dissociation energy of methane in the
gas-phase) and $\Delta r_i$ the change in bond length from the
equilibrium distance. $\gamma$ was
calculated by equating the second derivatives along one bond of the
harmonic and the Morse PES. 
If we transform Eq.\ (\ref{Vmorse_n}) back into normal mode
coordinates, we obtain
\begin{equation}
  \label{Vmorse_d}
  V_{\rm Morse}=D_e
  \sum_{i=1}^4\Big[1-e^{\gamma_{i2}X_2}e^{\gamma_{i3}X_3}e^{\gamma_{i4}X_4}
  e^{\gamma_{i7}X_7}e^{\gamma_{i8}X_8}e^{\gamma_{i9}X_9}
  e^{\gamma_{i,10}X_{10}}\Big]^2,
\end{equation}
with $D_e$ as above. $\gamma$'s are given in Tables \ref{tab:gammaone}
and \ref{tab:gammatwo}. Note that, although we have only changed the
PES of the bond lengths, the $\nu_4$ umbrella modes are also
affected. This is because these modes are not only bending, but also
contain some changes of bond length.

The new intramolecular PES now becomes
\begin{equation}
  \label{Vintra_morse}
  V_{\rm intra}=V_{\rm harm}+V_{\rm Morse}-V_{\rm corr}, 
\end{equation}
where $V_{\rm harm}$ is as given in Eq.\  (\ref{Vharm}) and $V_{\rm corr}$ is
the quadratic part of $V_{\rm Morse}$, which is already in $V_{\rm harm}$. 

\subsubsection{Intramolecular potential with weakening C--D bonds}
\label{sec:weak}

When the methane molecule approach the surface the overlap of substrate
orbitals and anti-bonding orbitals of the molecule weakens the C--D bonds.
We want to include this effect for the C--D bonds of the deuteriums
pointing towards the surface. We have redefined the $V_{\rm Morse}$ 
given in Eq.\ (\ref{Vmorse_d}) and also replace it in
Eq.\ (\ref{Vintra_morse}). A sigmoidal function is used to switch from the
gas phase C--D bond to a bond close to the surface. We have used the
following, somewhat arbitrary, approximations. 
(i) The point of inflection should be at a reasonable distance from the
surface. It is set to the turnaround point for a rigid methane molecule
with translation energy 93.2 kJ/mol plus twice the fall-off distance of
the interaction with the surface. (ii) The depth of the PES of the C--D
bond is 480 kJ/mol in the gas phase, but only 93.2 kJ/mol near the
surface. The value 93.2 kJ/mol corresponds to the height of the
activation barrier used in our dissociation.\cite{jan95} (iii) The
exponential factor is the same as for the interaction with the surface. 

If we transform to normal-mode coordinates for the particular
orientations, we then obtain
\begin{equation}
  \label{Vweak_d}
  V_{\rm weak}=D_e
  \sum_{i=1}^4W_i\Big[1-e^{\gamma_{i2}X_2}e^{\gamma_{i3}X_3}
  e^{\gamma_{i4}X_4}
  e^{\gamma_{i7}X_7}e^{\gamma_{i8}X_8}e^{\gamma_{i9}X_9}
  e^{\gamma_{i,10}X_{10}}\Big]^2,
\end{equation}
where $W_i=1$ for non-interacting bonds and
\begin{equation}
  \label{Wi}
  W_i= { {1 + \Omega e^{-\alpha_1X_1 + \omega}} \over {1 + 
  e^{-\alpha_1X_1 + \omega}} }  
\end{equation}
for the interacting bonds pointing towards the surface.
$\alpha_1$ is as given in Table \ref{tab:alpha}, $\gamma$'s are given in
Tables \ref{tab:gammaone} and \ref{tab:gammatwo}, $\Omega=1.942
\cdot10^{-1}$ and $\omega=7.197$.

\subsubsection{Intramolecular potential with elongation of the C--D bonds}
\label{sec:shift}

A weakened bond generally has not only a reduced bond strength, but also
an increased bond length.  We include the effect of the elongation of
the C--D bond length of the deuteriums that point towards the surface
due to interactions with the surface.  We have redefined the $V_{\rm
  Morse}$ given in Eq.\ (\ref{Vmorse_d}) and also replace it in Eq.\ 
(\ref{Vintra_morse}) for this type of PES.  We have used the following
approximations: (i) The transition state, as determined by
Ref.~\cite{bur93b} and \cite{bur93c}, has a C--H bond that is 0.54 {\AA}
longer than normal. This elongation should occur at the turn around
point for a rigid methane molecule with a translation energy of 93.2
kJ/mol.  (ii) The exponential factor is again the same as for the
interaction with the surface.

If we transform to normal-mode coordinates for the particular
orientations, then we obtain
\begin{equation}
  \label{Vshift_d}
  V_{\rm shift}=D_e
  \sum_{i=1}^4\Big[1-e^{\gamma_{i2}X_2}e^{\gamma_{i3}X_3}e^{\gamma_{i4}X_4}
  e^{\gamma_{i7}X_7}e^{\gamma_{i8}X_8}e^{\gamma_{i9}X_9}
  e^{\gamma_{i,10}X_{10}}\exp[S_ie^{-\alpha_1X_1}]\Big]^2 ,
\end{equation}
where $\alpha_1$ is as given in Table \ref{tab:alpha}, $\gamma$'s are
given in Tables \ref{tab:gammaone} and \ref{tab:gammatwo}. For
orientation with one deuterium towards the 
surface we obtain; $S_1=2.942 \cdot 10^{2}$ and $S_2=S_3=S_4=0$,
with two deuteriums; $S_1=S_2=0$ and $S_3=S_4=1.698 \cdot 10^{2}$,
and with three deuteriums; $S_1=0$ and $S_2=S_3=S_4=2.942 \cdot
10^{2}$.

\subsection{Initial States}
\label{sec:states}

The exact wave-function of a $D$-dimensional system, is expressed in the
MCTDH approximation by the form
\begin{equation}
\label{eD}
  \Psi_{\rm MCTDH}(q_1,\ldots,q_D;t)
  =\sum_{n_1\ldots n_D}
  c_{n_1\ldots n_D}(t)
  \,\psi_{n_1}^{(1)}(q_1;t)\ldots\psi_{n_D}^{(D)}(q_D;t).
\end{equation}
All initial states in the simulations start with the vibrational
ground state. The initial translational part $\psi^{({\rm tr})}$ is
represented by a Gaussian wave-packet, 
\begin{equation}
  \label{Gausswave}
  \psi^{({\rm tr})}(X_1)=(2\pi\sigma^2)^{-1/4}
  \exp\left[-{(X_1-X_0)^2\over 4\sigma^2}+
  iP_1X_1\right],
\end{equation}
where $\sigma$ is the width of the wave-packet (we used $\sigma=320.248$
atomic units), $X_0$ is the initial position (we used $X_0=11\sigma$,
which is far enough from the surface to observe no repulsion) and $P_1$
is the initial momentum. Since we used mass-weighted coordinates the
Gaussian wavepacket are identical for CD${}_4$ and CH${}_4$. 
We performed simulations in the energy range of
32 to 96 kJ/mol. We here present only the results of 96 kJ/mol
(equivalent to $P_1=-0.2704$ atomic units), because they showed the most
obvious excitation probabilities for $V_{\rm Morse}$. We used seven
natural single-particle states, 512 grid points and a grid-length of
$15\sigma$ for the translational coordinate. With this grid-width we can
perform simulation with a translational energy up to 144 kJ/mol.

Gauss-Hermite discrete-variable representations (DVR)\cite{light85}
were used to represent the wavepackets of the vibrational modes.
We used for all simulations of CD${}_4$ the same number of DVR points as
for CH${}_4$, which was 5 DVR points for the $\nu_2$ bending modes and 8
DVR points for the $\nu_4$ umbrella, $\nu_3$ asymmetrical
stretch, and $\nu_1$ symmetrical stretch mode for an numerical exact
integration, except for the simulations with $V_{\rm shift}$,
where we used 16 DVR points for the $\nu_1$ symmetrical stretch mode,
because of the change in the equilibrium position.

Also the same configurational basis was used for both isotopes.
We did the simulation with one bond pointing towards the
surface in eight dimensions, because the $\nu_2$ bending modes $X_5$ and
$X_6$ do not couple with the other modes. We needed four natural
single-particle states for modes $X_2$, $X_3$ and $X_4$, and just one
for the others. So the number of configurations was $7^1 \cdot 4^3 \cdot
1^4 = 448$.  
The simulation with two bonds pointing towards the surface was
performed in nine dimensions. One of the $\nu_2$ bending mode ($X_6$)
does not couple with the other modes, but for the other mode $X_5$ we
needed four natural single-particle states. The number of configurations was
$7^1 \cdot 4^4 \cdot 1^4 = 1792$, because we needed the same number of
natural single-particle states as mentioned above for the other modes.
We needed ten dimensions to perform the simulation with three bonds
pointing towards the surface. We used here one natural single-particle
state for the modes $X_5$ to $X_{10}$ and four natural single-particle
states for $X_2$ to $X_4$, which gave us $7^1 \cdot 4^3 \cdot 1^6 = 448$
configurations. 

\section{Results and Discussion}

\subsection{Excitation probabilities and structure deformation of CD${}_4$}

The scattering probabilities for CD${}_4$ are predominantly elastic,
as we also found in our previous simulations of CH${}_4$
scattering.\cite{mil98} The elastic scattering probability is larger than
0.99 for all orientation of the PESs with $V_{\rm Morse}$ and $V_{\rm
  weak}$ at a translational energy of 96 kJ/mol. For the PES with
$V_{\rm shift}$ we observe an elastic scattering probability of 0.981
for the orientation with one, 0.955 with two and 0.892 with three
deuteriums pointing towards the surface. This is lower than we have
found for CH${}_4$, which is 0.956 for the orientation with three
hydrogens pointing towards the surface and larger than 0.99 for the
others. The higher inelastic scattering probabilities of CD${}_4$ was
expected, because the force constants $k_i$ of CD${}_4$ are decreased up
to 50\% with respect to those of CH${}_4$ and the translational
surface repulsion fall-off differs only little.

When we look at the excitation probabilities at the surface for the PES
with $V_{\rm Morse}$ and $V_{\rm weak}$, then we observe generally an
increase for CD${}_4$ compared with CH${}_4$, except for the $\nu_4$
umbrella mode in the orientation with two bond pointing towards the
surface. Relevant differences in the structure deformations are observed
only in the bond angles, which are increased for CD${}_4$ in the
orientations with one and three bonds pointing towards the surface. The
bond angle deformation of the angle between the bonds pointing towards
the surface in the orientation with two bonds pointing towards the
surface is decreased for CD${}_4$. We observe again that the PES with
$V_{\rm weak}$ gives larger structure deformations than the PES with
$V_{\rm Morse}$, but the differences are smaller for CD${}_4$ than
CH${}_4$.

For the PES with $V_{\rm shift}$ we do not observe this effect on the
bond angle deformation. The bond angle deformation for the orientation
with two and three deuteriums pointing towards the surface is the same
as for CH${}_4$ and it is just $0.1^{\circ}$ less for the bond angle at
the surface side in the orientation with one deuterium pointing towards
the surface. The excitation probabilities (see Table \ref{tab:exprop})
for the $\nu_2$ bending and $\nu_4$ umbrella modes become higher for all
orientations for CD${}_4$, which is necessary for getting the same bond
angle deformations as CH${}_4$. 

The changes in the bond distances for the orientations with one and two
bonds pointing towards the surface is for CD${}_4$ almost the same as
for CH${}_4$. For the orientation with three bonds pointing towards the
surface, we found that the maximum bond lengthening of the bonds on the
surface side was $0.032$ {\AA} less for CD${}_4$ than CH${}_4$. We also
found that the bond shortening of the bond pointing away from the
surface is $0.010$ {\AA} more for CD${}_4$. These are only minimal
differences, which also only suggest that the bond deformation for CD${}_4$
has been influenced slightly more by the $\nu_3$ asymmetrical stretch
mode than the $\nu_1$ symmetrical stretch mode. The observed excitation
probabilities for these modes do not contradict this, but are not
reliable enough for hard conclusions because of their high magnitude.
It is also not clear, beside of this problem, what they really
represent. Is the excitation caused by a different equilibrium position
of the PES at the surface in a mode or is it caused by extra energy in this
mode? To answer these questions we decided to do an energy distribution
analysis during the scattering for both isotopes. 

\subsection{Energy distribution in CH${}_4$ and CD${}_4$}

The energy distribution analysis is performed by calculating the
expectation values of the important term of the Hamiltonian $H$ for the
wave-function $\Psi (t)$ at a certain time $t$ during the scattering of
CD${}_4$ and CH${}_4$ for all presented orientations in this and our
previous paper.\cite{mil98} We will present here only the results of the
PES with $V_{\rm shift}$, because it is the only model PES for which the
energy distribution analysis is relevant for the discussion of the
dissociation hypotheses later on.

We can obtain good information about the energy distribution per mode by
looking at the kinetic energy expectation values
$\langle\Psi (t)\vert T_j \vert\Psi (t)\rangle$ per mode $j$ (see Table
\ref{tab:kinch4}), because the kinetic energy
operators $T_j$ have no cross terms like the PESs have. When we discuss the
kinetic energy of a mode we normally refer to the $a_1$ mode of the 
C${}_{3v}$ or C${}_{2v}$ symmetry, because in these modes we have
observed the highest excitation probabilities and the change in kinetic
energy in the other modes is generally small.  

By looking at the expectation values of some terms of the PES
$\langle\Psi (t)\vert V_{\rm term} \vert\Psi (t)\rangle$ (see
Table \ref{tab:potch4}), we obtain information
about how the kinetics of the scattering is driven by the PES. 
The $V_{\rm surf}$ PES [see Eqs.\ (\ref{vsurfone}), (\ref{vsurftwo}) and
(\ref{vsurfthree})] is the surface hydrogen/deuterium 
repulsion for a given orientation. $V_{\rm harm}(\nu_2)$ and
$V_{\rm harm}(\nu_4)$ [see Eq.\ (\ref{Vharm})] are the pure harmonic
terms of the intramolecular 
PES of the $a_1$ modes in the C${}_{3v}$ and C${}_{2v}$ symmetry
corresponding to a $\nu_2$ bending and $\nu_4$ umbrella modes,
respectively. The pure harmonic correction terms of $V_{\rm corr}$ [see
Eq.\ (\ref{Vintra_morse})] are included in them. $V_{\rm bond}(R_{\rm up})$ and
$V_{\rm bond}(R_{\rm down})$ are the potential energy in a single C--H
or C--D bond pointing respectively towards and away from the surface,
and they give the expectation value of one bond term of $V_{\rm shift}$
[see Eq.\ (\ref{Vshift_d})].  
All given expectation values are the maximum deviation of the initial
values, which effectively means the values at the moment the molecule hits
the surface.

The largest changes in expectation values are, of course, in
the kinetic energy of the translational mode. The translational kinetic
energy does not become zero as we should expect in classical dynamics.
The loss of translational kinetic energy is primary absorbed by the
$V_{\rm surf}$ terms of the PESs. The expectation values of the $V_{\rm
  surf}$ terms show  the ability of the hydrogens or deuteriums to come
close to the metal surface, since in real space their exponential
fall-offs are the same for both isotopes. For a rigid molecule the sum
of the translational kinetic energy and $V_{\rm surf}$ should be
constant, so all deviations of this sum have to be found back in the
intramolecular kinetic energy and other PES terms. 

We observe that both the minimum in the translational kinetic energy and
the maximum in the $V_{\rm surf}$ terms were higher for CH${}_4$ than
CD${}_4$, so we have to find more increase in energy in the
intramolecular modes and PES terms for CD${}_4$ than CH${}_4$. We indeed
do so and that can be one of the reasons we found higher inelastic
scatter probabilities for CD${}_4$ for the PES with $V_{\rm shift}$ .  

For the orientations with one and two bonds pointing towards the surface
we observe a large increase of the kinetic energy in the $\nu_3$
asymmetrical stretch mode. If we compare this with the excitation
probabilities, we find that the kinetic energy analysis gives indeed a
different view on the dynamics. For the orientation with two bond
pointing towards the surface we have found for both isotopes very high
excitation probabilities in the $\nu_1$ and $\nu_3$ stretch modes. We
know now from the kinetic energy distribution that for the $\nu_1$
symmetrical stretch mode the high excitation probability is caused by
the change of the equilibrium position of the $\nu_1$ mode in the PES
and that for the $\nu_3$ stretch mode probably the PES also has become
narrower. 

For the orientation with three bonds pointing towards the surface we also
obtain an large increase of the kinetic energy of the $\nu_3$
asymmetrical stretch mode, but we also find an even larger increase in
the kinetic energy of the $\nu_1$ symmetrical stretch mode. The total
kinetic energy was extremely large, because the kinetic energy of the
translational mode becomes also much larger than for 
the other orientations. Because of this the $V_{\rm surf}$ terms had to
be around twice as low as for the other orientations. 

All $V_{\rm bond}(R_{\rm up})$ terms 
become lower compared to the initial value, especially in the
orientation with two bond pointing towards the surface. In the
orientation with one bond pointing towards the surface, the $V_{\rm
  bond}(R_{\rm down})$ term became higher. This is caused by the
repulsion of $V_{\rm surf}$ in the direction of the bond. The
increase of this PES term value is higher for CD${}_4$ than CH${}_4$.

In the orientation with three bond pointing towards the surface we also
observe a higher $V_{\rm bond}(R_{\rm down})$ value, with also the
highest increase for CD${}_4$. In relation with the somewhat shorter
bond distance for the $R_{\rm down}$ of CD${}_4$ compared with
CH${}_4$, which was also a bit lower compared with the other
orientations, we know now that the hydrogens and especially the
deuterium have problems in following the minimum energy path of the PES
with $V_{\rm shift}$ during the scattering dynamics. This leads to higher
kinetic energy in the vibrational modes, which results in more inelastic
scattering. 

The $V_{\rm harm}(\nu_2)$ term 
increases in respect to the initial value, but not as much as the
increase of the $V_{\rm harm}(\nu_4)$ is for the orientation with two
bonds pointing towards the surface. The values of $V_{\rm harm}(\nu_4)$
for CD${}_4$ are even higher than for CH${}_4$. We observe also a larger
increase of the kinetic energy in the $\nu_4$ umbrella mode for CD${}_4$
than for CH${}_4$. So although there is somewhat more energy transfer to
the vibrational modes for CD${}_4$ than CH${}_4$, this extra vibrational
energy is absorbed especially in the $\nu_4$ umbrella mode of CD${}_4$.

\subsection{Dissociation hypotheses}

We like now to discuss some possible implications of the scattering
simulation for the isotope effect on the dissociation of methane.  In
our previous paper we have already drawn some conclusions about the
possible reaction mechanism and which potential type would be necessary
for dissociation.\cite{mil98} We found the direct breaking of a single
C--H bond in the initial collision more reasonable than the splats model
with single bond breaking after an intermediary Ni--C bond formation as
suggested by Ref.~\cite{lee87}, because the bond angle deformations
seems to small to allow a Ni--C to form. From the simulations with
CD${}_4$ we can draw the same conclusions. The PES with $V_{\rm shift}$
gives the same angle deformations for both isotopes, which is not
sufficient for the splats model. The other potentials give higher bond
angle deformations for the orientation with three deuteriums pointing
towards the surface. If the Ni--C bond formation would go along this
reaction path, then the dissociation of CD${}_4$ should be even more
preferable than CH${}_4$, which is not the case.  So we only have to
discuss the implication of the scattering simulation for the
dissociation probabilities of CH${}_4$ and CD${}_4$ for a direct
breaking of a single bond reaction mechanism.  This reaction mechanism
can be influenced by what we will call a direct or an indirect effect.

A direct effects is the expected changes in the dissociation probability
between CH${}_4$ and CD${}_4$ for a given orientation. Since we expect
that we need for dissociation a PES with an elongation of the bonds
pointing towards the surface, we only have to look at the isotope effect
in the simulation for the PES with $V_{\rm shift}$ for different
orientations to discuss some direct effect. It is clear from our
simulations that the bond lengthening of CD${}_4$ is smaller than
CH${}_4$ for the orientation with three bonds pointing towards the
surface. If this orientation has a high contribution to the dissociation
of methane, then this can be the reason of the higher dissociation
probability of CH${}_4$. In this case our simulations also explain why
Ref.\cite{car98} did not observe a high enough isotope effect in
the dissociation probability of their simulation with CH${}_4$ and CD${}_4$
modelled by a diatomic, because we do not observe a change in bond
lengthening between the isotopes for the orientation with one bond
pointing towards the surface.

The orientation with three bonds pointing towards the surface is also
the orientation with the highest increase of the total vibrational
kinetic energy for the PES with $V_{\rm shift}$, because the energy
distribution analysis shows besides an high increase of the kinetic
energy in the $\nu_3$ asymmetrical stretch mode also an high increase in
the $\nu_1$ symmetrical stretch mode. Since vibrational kinetic energy
can be used effectively to overcome the dissociation barrier, the
orientation with three bonds indicates to be a more preferable orientation
for dissociation. Moreover the relative difference in kinetic energy
between both isotopes is for the $\nu_1$ stretch mode larger than
for the $\nu_3$ stretch mode. If the kinetic energy in the $\nu_1$
stretch mode contributes significantly to overcoming the dissociation
barrier, then it is another explanation for the low isotope effect in
Ref.\cite{car98}. 

An indirect effect is the expected changes in the dissociation
probability between CH${}_4$ and CD${}_4$ through changes in the
orientations distribution caused by the isotope effect in the
vibrational modes. This can be the case if the favourable
orientation for dissociation is not near the orientation with three
bonds pointing towards the surface, but more in a region where one or
two bonds pointing towards the surface. These orientations do not show a
large difference in deformation for the PES with $V_{\rm shift}$. We can
not draw immediate conclusion about the indirect effect from our
simulations, since we did not include rotational motion, but our
simulation show that an indirect isotope effect can exist. For the PES
with $V_{\rm Morse}$ in the orientation with three bonds pointing
towards the surface, we observe that CD${}_4$ is able to come closer to
the surface than CH${}_4$. So this rotational orientation should be more
preferable for CD${}_4$ than for CH${}_4$. On the other hand, if the PES
is in this orientation more like $V_{\rm shift}$ the dissociation
probability in other orientation can be decreased for CD${}_4$ through
higher probability in inelastic scattering channels. 

So for both effects the behaviour of the orientation with three bonds
pointing towards the surface seems to be essential for a reasonable
description of the dissociation mechanism of methane.  A wavepacket
simulation of methane scattering including one or more rotational
degrees of freedom and the vibrational stretch modes will be a good
starting model to study the direct and indirect effects, since most of
the kinetic energy changes are observed in the stretch modes and so the
bending and umbrella modes are only relevant with accurate PESs.
Eventually dissociation paths can be introduced in the PES along one or
more bonds. 

Beside of our descriptions of the possible isotopes effect for
the dissociation extracted of the scatter simulations we have to keep
in mind that also a tunneling mechanism can be highly responsible for the
higher observed isotope effect in the experiment and that a different
dissociation barrier in the simulations can enhance this effect of tunneling. 

\section{Conclusions}

The scattering is in all cases predominantly elastic. However, the
observed inelastic scattering is higher for CD${}_4$ compared with
previous simulation on CH${}_4$ for the PES with an elongated
equilibrium bond length close to the surface.  When the molecule hits
the surface, we observe in general a higher vibrational excitation for
CD${}_4$ than CH${}_4$. The PES with an elongated equilibrium bond
length close to the surface gives for both isotopes almost the same
deformations, although we observe a somewhat smaller bond lengthening
for CD${}_4$ in the orientation with three bonds pointing towards the
surface.  The other model PESs show differences in the bond angle
deformations and in the distribution of the excitation probabilities of
CD${}_4$ and CH${}_4$, especially for the PES with only an anharmonic
intramolecular potential.

Energy distribution analysis contributes new information on the
scattering dynamics.  A higher transfer of translational energy towards
vibrational kinetic energy at the surface results in higher inelastic
scattering. The highest increase of vibrational kinetic energy is found
in the $\nu_3$ asymmetrical stretch modes for all orientations and also
in the $\nu_1$ symmetrical stretch mode for the orientation with three
bonds pointing towards the surface, when the PES has an elongated
equilibrium bond length close to the surface.

Our simulations give an indication that the isotope effect in the
methane dissociation is caused mostly by the difference in the
scattering behaviour of the molecule in the orientation with three bonds
pointing towards the surface. At least multiple vibrational stretch
modes should be included for a reasonable description of isotope effect
in the methane dissociation reaction. 

\section*{Acknowledgments}
This research has been financially supported by the Council for Chemical
Sciences of the Netherlands Organization for Scientific Research (CW-NWO). 
This work has been performed under the auspices of NIOK, the Netherlands
Institute for Catalysis Research, Lab Report No. TUE-99-5-02.

\begin{table}[H]
  \caption{Overview of the relations between the mass-weighted coordinates
  $X_i$; the force constants $k_i$ (in atomic units) for CD${}_4$, the
  designation, and the symmetry in T${}_d$, C${}_{3v}$ and
  C${}_{2v}$.} 
  \label{tab:gen}
  \begin{tabular}{r l l c c c}
    \hline
    $i$ & $k_i$ & designation & T${}_d$ & C${}_{3v}$ & C${}_{2v}$ \\
    \hline
    $1$ &                      & translation & $t_2$ & $a_1$ & $a_1$ \\ 
    $2$ & $8.897\cdot 10^{-5}$ & $\nu_1$; symmetrical stretch & $a_1$ &
    $a_1$  & $a_1$ \\ 
    $3$ & $2.008\cdot 10^{-5}$ & $\nu_4$; umbrella & $t_2$ & $a_1$ & $a_1$ \\ 
    $4$ & $1.060\cdot 10^{-4}$ & $\nu_3$; asymmetrical stretch & $t_2$ &
       $a_1$ & $a_1$ \\  
    $5$ & $2.447\cdot 10^{-5}$ & $\nu_2$; bending & $e$ & $e$ & $a_1$ \\
    $6$ & $2.447\cdot 10^{-5}$ & $\nu_2$; bending & $e$ & $e$ & $a_2$ \\
    $7$ & $2.008\cdot 10^{-5}$ & $\nu_4$; umbrella & $t_2$ & $e$ & $b_1$ \\ 
    $8$ & $2.008\cdot 10^{-5}$ & $\nu_4$; umbrella & $t_2$ & $e$ & $b_2$ \\  
    $9$ & $1.060\cdot 10^{-4}$ & $\nu_3$; asymmetrical stretch & $t_2$ &
       $e$ & $b_1$ \\  
    $10$ & $1.060\cdot 10^{-4}$& $\nu_3$; asymmetrical stretch & $t_2$ &
       $e$ & $b_2$ \\  
    \hline
  \end{tabular}
\end{table}

\begin{table}[H]
  \caption{$\alpha$ and $\beta$ values (in atomic units) of $V_{\rm
    surf}$ for CD${}_4$ with one, two or three deuteriums pointing
    towards the surface 
    [see Eqs.\  (\ref{vsurfone}), (\ref{vsurftwo}) and (\ref{vsurfthree})].} 
  \label{tab:alpha}
  \begin{tabular}{l r r r}
    \hline
    & one & two & three \\
    \hline
    $\alpha_1$ & $5.617\cdot 10^{-3}$  & $5.617\cdot 10^{-3}$  &
    $5.617\cdot 10^{-3}$ \\ 
    $\alpha_2$ & $8.882\cdot 10^{-3}$  & $5.128\cdot 10^{-3}$  &
    $2.960\cdot 10^{-3}$ \\
    $\alpha_3$ & $4.703\cdot 10^{-3}$  & $-4.614\cdot 10^{-3}$ &
    $-7.720\cdot 10^{-3}$ \\
    $\alpha_4$ & $-1.353\cdot 10^{-2}$ & $-5.103\cdot 10^{-3}$ &
    $-2.295\cdot 10^{-3}$ \\
    $\alpha_5$ &                       & $-7.252\cdot 10^{-3}$  &
    \\
    $\beta_{1}$ & &                          & $4.187\cdot 10^{-3}$ \\
    $\beta_{2}$ & &                          & $7.252\cdot 10^{-3}$ \\
    $\beta_{3}$ & & $4.659\cdot 10^{-3}$ & $2.196\cdot 10^{-3}$ \\
    $\beta_{4}$ & &                      & $3.804\cdot 10^{-3}$ \\
    $\beta_{5}$ & & $4.212\cdot 10^{-3}$ & $2.295\cdot 10^{-3}$ \\
    $\beta_{6}$ & &                      & $3.439\cdot 10^{-3}$ \\
    \hline
  \end{tabular}
\end{table}

\begin{table}[H]
  \caption{$\gamma$ values (in atomic units) of $V_{\rm Morse}$ for
    CD${}_4$ with one and three deuteriums pointing towards the surface
    [see Eq.\  (\ref{Vmorse_d})].} 
  \label{tab:gammaone}
  \begin{tabular}{l l r}
    \hline
    one & three & value\\
    \hline
    $\gamma_{12},\gamma_{22}, \gamma_{32}, \gamma_{42}$ &
    $\gamma_{12}, \gamma_{22}, \gamma_{32}, \gamma_{42}$ &
    $7.629\cdot 10^{-3}$ \\ 
    $\gamma_{13}, -3\gamma_{23}, -3\gamma_{33}, -3\gamma_{43}$ & 
    $-\gamma_{13}, 3\gamma_{23}, 3\gamma_{33}, 3\gamma_{43}$ &
    $1.397\cdot 10^{-3}$ \\
    $\gamma_{14}, -3\gamma_{24}, -3\gamma_{34}, -3\gamma_{44}$ &
    $-\gamma_{14}, 3\gamma_{24}, 3\gamma_{34}, 3\gamma_{44}$  &
    $-1.454 \cdot 10^{-2}$ \\
    $\gamma_{17}, \gamma_{18}, \gamma_{19}, \gamma_{1,10}, \gamma_{28}, 
    \gamma_{2,10}$ & 
    $\gamma_{17}, \gamma_{18}, \gamma_{19}, \gamma_{1,10}, \gamma_{28},  
    \gamma_{2,10}$  & $0.0$ \\
    $\gamma_{27}, -2\gamma_{37}, -2\gamma_{47}$ &
    $-\gamma_{27}, 2\gamma_{37}, 2\gamma_{47}$ & $1.318 \cdot 10^{-3}$
    \\
    $\gamma_{38}, -\gamma_{48}$ & $\gamma_{38}, -\gamma_{48}$ &
    $-1.114 \cdot 10^{-3}$ \\ 
    $\gamma_{29}, -2\gamma_{39}, -2\gamma_{49}$ &
    $-\gamma_{29}, 2\gamma_{39}, 2\gamma_{49}$ & $-1.371 \cdot
    10^{-2}$ \\
    $\gamma_{3,10}, -\gamma_{4,10}$ & $-\gamma_{3,10}, \gamma_{4,10}$ &
    $1.187 \cdot 10^{-2}$ \\  
    \hline
  \end{tabular}
\end{table}

\begin{table}[H]
  \caption{$\gamma$ values (in atomic units) of $V_{\rm Morse}$ for
    CD${}_4$ with two deuteriums pointing towards the surface [see Eq.\
    (\ref{Vmorse_d})].}   
  \label{tab:gammatwo}
  \begin{tabular}{l  r}
    \hline
    two & value\\
    \hline
    $\gamma_{12}, \gamma_{22}, \gamma_{32}, \gamma_{42}$ & $7.629\cdot
    10^{-3}$ \\ 
    $\gamma_{13}, \gamma_{23}, -\gamma_{33}, -\gamma_{43}, \gamma_{17},
    -\gamma_{27}, \gamma_{37}, -\gamma_{47}, \gamma_{18}, -\gamma_{28},
    -\gamma_{38}, \gamma_{48}$ & $-8.070 \cdot 10^{-4}$  \\
    $\gamma_{14}, \gamma_{24}, -\gamma_{34}, -\gamma_{44}, \gamma_{19},
    -\gamma_{29}, \gamma_{39}, -\gamma_{49}, \gamma_{1,10},
    -\gamma_{2,10}, -\gamma_{3,10}, \gamma_{4,10}$ & $8.396 \cdot
    10^{-3}$ \\ 
    \hline
  \end{tabular}
\end{table}

\begin{table}[H]
  \caption{Excitation probabilities at the surface, at an initial
    translational energy of 96 kJ/mol and all initial vibrational states
    in the ground state, for the intramolecular PES with elongation of
    the C--D bonds [see Eq.\ (\ref{Vshift_d})] in the $a_1$ modes of the
    C${}_{3v}$ and C${}_{2v}$ symmetry, with one, two or three
    deuteriums pointing towards the surface.  These modes are a
    $\nu_1(a_1)$ symmetrical stretch, a $\nu_2(e)$ bending, a
    $\nu_3(t_2)$ asymmetrical stretch, and a $\nu_4(t_2)$ umbrella. In
    parenthesis is the irreducible representation in T${}_d$ symmetry.}
  \label{tab:exprop}
  \begin{tabular}[t]{l c c c c}
    \hline
      orientation & $\nu_1(a_1)$ stretch & $\nu_2(e)$ bending &
      $\nu_3(t_2)$ stretch & $\nu_4(t_2)$ umbrella \\
    \hline
      one   & 0.460 &       & 0.910 & 0.174 \\
      two   & 0.792 & 0.092 & 0.830 & 0.495 \\
      three & 0.868 &       & 0.756 & 0.387  \\
    \hline
  \end{tabular}
\end{table}

\begin{table}[H]
  \caption{Expectation values of the kinetic energy per mode in
    mHartree for CH${}_4$ and CD${}_4$, at an initial translational energy 
    of 96 kJ/mol and all initial vibrational states in the ground state,
    for the intramolecular PES with elongation of
    the C--H/D bonds [see Eq.\ (\ref{Vshift_d})] in
    the $a_1$ modes of the C${}_{3v}$  and C${}_{2v}$ symmetry, with
    one, two or three deuteriums pointing towards the surface. 
    These modes are a $\nu_1(a_1)$ symmetrical stretch, a $\nu_2(e)$
    bending , a $\nu_3(t_2)$ asymmetrical stretch, and a $\nu_4(t_2)$
    umbrella. In parenthesis is the irreducible representation in 
    T${}_d$ symmetry.}
  \label{tab:kinch4}
  \begin{tabular}[t]{l l c c c c c}
    \hline
      isotope & orientation &  translation & $\nu_1(a_1)$ stretch &
      $\nu_2(e)$ bending & 
      $\nu_3(t_2)$ stretch & $\nu_4(t_2)$ umbrella \\
    \hline
    CH${}_4$ & initial & 36.57 & 3.30 & 1.75 & 3.39 & 1.50 \\
             & one     & 16.76 & 3.56 &      & 4.53 & 1.51 \\
             & two     & 14.59 & 3.50 & 1.79 & 4.67 & 1.57 \\
             & three   & 20.53 & 5.32 &      & 4.39 & 1.58 \\
    \hline
    CD${}_4$ & initial & 36.57 & 2.33 & 1.24 & 2.52 & 1.12 \\
             & one     & 16.17 & 2.61 &      & 4.09 & 1.18 \\
             & two     & 14.00 & 2.78 & 1.27 & 4.05 & 1.27 \\
             & three   & 20.06 & 4.37 &      & 3.80 & 1.28 \\
    \hline
  \end{tabular}
\end{table}

\begin{table}[H]
  \caption{Expectation values of the potential energy terms in
    mHartree for CH${}_4$ and CD${}_4$, at an initial translational energy 
    of 96 kJ/mol and all initial vibrational states in the ground state,
    for the intramolecular PES with elongation of
    the C--H/D bonds [see Eq.\ (\ref{Vshift_d})].
    $V_{\rm surf}$ is the total surface hydrogen repulsion; $V_{\rm
    harm}(\nu_2)$ and $V_{\rm harm}(\nu_4)$ are the harmonic
    terms of the intramolecular PES of the $a_1$ modes in the C${}_{3v}$
    and C${}_{2v}$ symmetry corresponding to a $\nu_2(e)$ bending and
    $\nu_4(t_2)$ umbrella modes respectively in the T${}_d$ symmetry.
    $V_{\rm bond}(R_{\rm up})$ and $V_{\rm bond}(R_{\rm down})$ are the
    potential energy in a single C--H/D bond pointing respectively towards
    and away from the surface.} 
  \label{tab:potch4}
  \begin{tabular}[t]{l l c c c c c}
    \hline
      isotope & orientation &  $V_{\rm surf}$ & $V_{\rm harm}(\nu_2)$ & 
       $V_{\rm harm}(\nu_4)$ & $V_{\rm bond}(R_{\rm up})$ &
      $V_{\rm bond}(R_{\rm down})$ \\  
    \hline
    CH${}_4$ & initial &  0.00 & 1.75 & 1.50 & 3.39 & 3.39 \\
             & one     & 18.20 &      & 1.87 & 3.25 & 3.85 \\
             & two     & 18.55 & 2.18 & 4.01 & 2.75 & 3.45 \\
             & three   &  9.22 &      & 2.94 & 3.00 & 3.74 \\
    \hline
    CD${}_4$ & initial &  0.00 & 1.24 & 1.12 & 2.48 & 2.48 \\
             & one     & 17.94 &      & 1.89 & 2.43 & 3.44 \\
             & two     & 18.45 & 1.68 & 4.52 & 2.29 & 2.74 \\
             & three   &  8.71 &      & 3.49 & 2.28 & 3.21 \\
    \hline
  \end{tabular}
\end{table}

\end{document}